# Effective Delegation and Leadership in Software Management

Star Dawood Mirkhan[1], Skala Kamaran Omer[2], Hussein Mohammed Ali[3], Mahmood Yashar Hamza[4], Tarik Ahmed Rashid[5], and Poornima Nedunchezhian[6]

[1]Computer Science and Engineering, University of Kurdistan Hewler, Kurdistan Region, Iraq - E-mail: satar.dawood@ukh.edu.krd

[2]Computer Science and Engineering, University of Kurdistan Hewler, Kurdistan Region, Iraq - E-mail: skala.kamaran@ukh.edu.krd

[3]Computer Engineering Department, Tishk International University, Kurdistan Region, Iraq - E-mail: hussein.mohammed@tiu.edu.iq

[4]Computer Engineering Department, Tishk International University, Kurdistan Region, Iraq - E-mail: mahmood.yashar@tiu.edu.iq

[5]Computer Science and Engineering, University of Kurdistan Hewler, Kurdistan Region, Iraq - E-mail: tarik.ahmed@ukh.edu.krd

[6]Computer Science and Engineering, Vellore Institute of Technology, Vellore, India – E-mail: dr.poornimanedunchezhian@gmail.com

* Corresponding author: Computer Science and Engineering, University of Kurdistan Hewler, Kurdistan Region, Iraq. Tel: +9647507330321

**Abstract**

**Delegation and leadership are critical components of software management, as they play a crucial role in determining the success of the software development process. This study examined the relationship between delegation and leadership in software management and the impact of these factors on project outcomes. Results showed that effective delegation and transformational leadership styles can improve workflow, enhance team motivation and productivity, and ultimately lead to successful software development projects. The findings of this study have important implications for software management practices, as they suggest that organizations and software managers should prioritize the development of effective delegation and leadership practices to ensure the success of their software development initiatives. Further research is needed to explore the complex interplay between delegation and leadership in software management and to identify best practices for improving these processes.**

## 1. Introduction

Effective management of software projects is crucial, in today's organizations as it encompasses the development, implementation, and maintenance of software systems. A strong leader with the ability to delegate tasks efficiently plays a role in ensuring software management (Weerasinghe , 2022). This paper investigates the interplay between delegation and leadership within the context of software management. How these behaviors impact software development outcomes. The study has two objectives, first, to analyze how delegation and leadership practices influence software development results, second to identify the factors that shape these practices and strategies employed by software managers to lead their teams effectively and delegate tasks (Zhang, et al., 2009). The structure of the paper is organized as follows: Section 2 provides a background on the importance of delegation and leadership in the paced realm of software management, Section 3 presents a review of relevant studies and existing literature related to the topic, Section 4 explores various theories underpinning delegation and leadership establishing a theoretical foundation for our research, in Section 5 delves into different types of leadership styles and delegation techniques considering their applicability, within software management contexts, in section 6 presents the techniques and methods on how to enhance the delegation and leadership, in Section 7 presents the difficulties that arise when delegating responsibilities and leading in the software industry. It also sheds light on strategies and recommendations to successfully overcome these challenges. Section 8 goes into the individuals who benefit from delegation in the software industry and explores the impact of delegation and leadership, on software management. The conclusion of the paper is in section 9, which presents the findings of our study, highlights takeaways, and suggests research avenues and it will wrap up by discussing how our research contributes to theory and practice. And the last section is the future work and studies.

## 2. Background

In the evolving world of software management, where constant innovation and quick development require strong leadership and effective delegation, which become crucial factors. Software projects are known for their complexity demanding skills and expertise to navigate through the intricacies of design, development, testing, and maintenance (F.A.Z & R.D.F.S.M, 2020). As technology advances at a pace software management continuously faces challenges to adjust and thrive. Also, the types of leadership and their characteristics are shown in Table 1

(Li, Tan, & Teo, 2012).

Table 1. Types of Leadership and Their Characteristics.

| Leadership Style | Characteristics |
|---|---|
| Transformational | Inspires, motivates, and sets a clear vision |
| Autocratic | Makes decisions unilaterally |
| Democratic | Involves the team in decision-making |
| Laissez-faire | Offers limited guidance |
| Servant Leadership | Focuses on serving the needs of others |

2.1 The Dynamic Nature of Software Management

Managing software involves overseeing the conception, creation, and maintenance of software systems that drive industries and applications. Whether it's developing cutting-edge apps, handling enterprise software, or building embedded systems for emerging technologies software management is a dynamic and diverse field. In this changing environment, successful software projects require not only technical expertise but also effective leadership and delegation. Given the evolution of technology, software managers must be flexible and responsive, to evolving client needs emerging development methods, and shifting market dynamics (Clark, Jones, & Armstrong, 2007), also represented in Table 2.

Table 2. Steps in Effective Delegation.

| Steps in Effective Delegation |
|---|
| Task Identification |
| Team Member Selection |
| Clear Communication |
| Support and Guidance |
| Performance Assessment |

2.2 The Role of Delegation

In the realm of software management, delegation is a method, for distributing tasks that aims to maximize project results. By assigning responsibilities to team members or subordinates' managers can leverage their strengths and expertise, leading to a more streamlined workflow. Delegation empowers managers to concentrate on choices, such, as defining project objectives allocating resources, and shaping the overall course of software development efforts.

Effective delegation plays a role, in software management. It involves steps that managers need to follow. First, they must carefully identify tasks that can be delegated, considering the capabilities and expertise of their team members. Selecting the individuals for these tasks is equally important requiring an understanding of their skills and experiences. Communication is key throughout the process with precise articulation of task objectives, deadlines, and expected outcomes. Managers also need to establish expectations regarding quality, timelines, and deliverables. Providing support and guidance is essential as tasks progress, including addressing any questions or concerns that arise offering feedback, and resolving challenges along the way. Regular performance assessments allow managers to provide feedback on team members' progress and gain an understanding of their performance while identifying areas, for improvement (Goericke, 2020).

2.3 The Significance of Leadership

Effective software management leadership extends beyond assigning tasks, it involves motivating and directing teams towards shared objectives. Leaders establish the path that offers guidance and foster a nurturing environment that empowers teams to accomplish project goals and improve their performance. Table 3 illustrates some of the attributes of effective leadership (Popa, Mihele, Făgărășan, & Pîslă, 2021).

**Table** 3. Attributes of Effective Leadership.

| Leadership Attributes |
|---|
| Motivation and Inspiration |
| Effective Communication |
| Strategic Decision-Making |
| Supportive Environment |

In the realm of software management, effective leadership involves several key attributes:

- **Motivation and Inspiration**: Leaders motivate and inspire team members, instilling a sense of purpose and commitment.
- **Effective Communication**: Communication is paramount, ensuring that project objectives and expectations are conveyed.
- **Strategic Decision-Making**: Leaders make strategic decisions that drive project success and organizational growth.

Transformational leadership has shown its effectiveness in software management. This style of leadership emphasizes inspiration, motivation, and intellectual stimulation, aligning team members with a shared vision and purpose (Peesker, Ryals, Rich, & Boehnke, 2019).

2.4 The Interplay Between Delegation and Leadership

Delegation and leadership go hand in hand in software management working together to create a relationship for an ideal software project environment (Weerasinghe , 2022). Effective delegation allows leaders to focus on their role of guiding and motivating teams while strong leadership fosters an environment of trust, motivation, and collaboration that enables delegation. The importance of this interplay, between delegation and leadership cannot be overstated within the context of software management. They contribute significantly to improved efficiency, team motivation, project success, and overall organizational growth. However, the success of this interplay also relies on the culture which can either support or hinder the implementation of these practices (Li, Tan, & Teo, 2012).

In the paced world of software management, it is crucial to understand how delegation and leadership work together synergistically to drive project outcomes and achieve success. This paper aims to delve into these dynamics by exploring their influence, challenges, best practices, and implications within the realm of software management. By shedding light on the role played by delegation and leadership it seeks to provide insights and guidance, for organizations striving to thrive in this ever-evolving landscape (Zhang, et al., 2009).

**3. Literature Review**

In this section, we present an examination of the literature and research regarding the relationship, between delegation and leadership in the software management domain. Our review of existing studies aims to establish an understanding of the importance of these concepts their influence on project results and the current state of research, in this area.

## 3.1 Delegation in Software Management

Delegation, an element of software management has attracted the interest of both researchers and professionals. In the initial research conducted by (Baird & Maruping, 2021), they emphasize the significance of delegation, in assigning tasks and optimizing software development projects. Delegation enables managers to assign responsibilities to team members leveraging their expertise and skills to enhance project efficiency and quality. Effective communication has been identified as an aspect of delegation according (Morrison-Smith & Ruiz, 2020). When team members understand their assigned tasks, objectives, and deadlines it minimizes the chances of misunderstandings and delays, in project completion. (Al-Saqqa, Sawalha, & AbdelNabi, 2020) emphasize the significance of providing support and guidance to individuals who are delegated tasks. Effective software managers do not assign responsibilities. Also, ensure that team members have the necessary resources and assistance to fulfill their duties efficiently.

## 3.2 Leadership in Software Management

Being a leader, in software management involves more than assigning tasks. It entails the capacity to inspire, guide, and create an environment that promotes team success. Valuable findings from the research indicate that transformational leadership, as suggested by (Popa, Mihele, Făgărășan, & Pîslă, 2021) has gained recognition in the field of software management. Transformational leaders can inspire and motivate their team members instilling a sense of purpose and dedication, towards project objectives. (Spiegler, Heinecke, & Wagner, 2021) highlight the importance of leadership, in adapting to emerging technologies and promoting innovation within software development teams. Good leaders foster. Offer support for teams to explore new and inventive solutions. The impact of culture on the effectiveness of leadership, in software management has been extensively studied. Organizational culture can impede the implementation of leadership approaches and delegation strategies (Goericke, 2020).

## 3.3 Interplay Between Delegation and Leadership

The literature also delves into the interconnectedness of delegation and leadership, in software management. Research has shown that effective delegation is often supported by leadership styles as they foster an environment that promotes trust, motivation, and collaboration (Spiegler, Heinecke, & Wagner, 2021). The synergy between delegation and leadership plays a role in attaining success, in projects.

## 3.4 Current Gaps in the Literature

Although the current body of literature offers perspectives, on software management there are still some areas that require attention. The existing research tends to concentrate on aspects like personality traits and skills when discussing delegation and leadership often neglecting the influence of organizational culture, team dynamics, and technology. Furthermore, there is a lack of studies exploring the lasting effects of effective delegation and leadership indicating the need, for further investigation (Athey, 1998). The sections will expand on this literature review to explore how delegation and leadership intertwine in software management, it will also examine their effects, on project outcomes, team performance, and employee satisfaction.

## 4. Delegation and Leadership Theories

These theories encompass aspects of delegation and leadership within the realm of software management. Trait theory focuses on the qualities that effective leaders possess, such as intelligence and charisma. Transformational leadership revolves around leaders' ability to inspire and motivate teams, towards shared goals. Transactional leadership involves using rewards and consequences to drive performance. Situational leadership suggests that leadership styles should adapt to team maturity and specific situations. Lastly, the path-goal theory highlights the role of leaders in clarifying paths to success and providing support. Together these theories guide our exploration of delegation and leadership, which allows us to understand their dynamics deeply (Zhang, et al., 2009).

## 5. Types of Leadership and Delegation

This section will delve into leadership and delegation styles. Discuss their importance and suitability, in the realm of software management. Each style possesses traits and consequences, for software development projects.

### 5.1 Leadership Styles

Different leadership styles have an impact, on how teams function make decisions, and achieve project goals. The following Table 4 provides a summary of leadership styles and their important characteristics:

Table 4. Leadership Styles.

| Leadership Style | Description |
|---|---|
| Autocratic Leadership | The leader makes decisions unilaterally with minimal team input. |
| Transformational Leadership | Inspires and motivates teams towards shared goals. |
| Servant Leadership | Focuses on serving the needs of team members and their growth. |
| Laissez-faire Leadership | Provides autonomy to the team with minimal guidance. |

There are leadership styles that can be applied in situations. For instance, when there's a need, for decision making an autocratic leadership style may be appropriate. On the other hand, if the goal is to encourage team innovation a transformational leadership style can be beneficial. Lastly when promoting autonomy, among team members is important a laissez-faire leadership style can be effective. Software managers have the opportunity to select a leadership style that aligns with their project goals and the dynamics of their team (Wong, 2003).

5.2 Delegation Styles

Delegation plays a role, in the management of software. In Table 5 you'll find an overview of delegation styles along, with their specific characteristics:

Table 5. Delegation Styles.

| Delegation Style | Description |
|---|---|
| Directive Delegation | The delegate is provided with specific instructions on task execution. |
| Supportive Delegation | The delegate is given resources and support to perform tasks. |
| Empowerment Delegation | The delegate is granted authority to make decisions independently. |
| Delegation by Objectives | The delegate receives specific objectives and autonomy for execution. |

Different delegation styles have purposes ranging from providing instructions, for tasks (directive) to fostering initiative (empowerment). When it comes to delegation software managers must choose the style that aligns with the nature of the tasks at hand and the abilities of their team members. By understanding these styles, software managers can make informed decisions that contribute to team performance and successful project outcomes. The tables presented offer an overview of each style's characteristics, which can assist software managers in selecting the suitable approach, for their specific software development projects (Klein, Ziegert, Knight, & Xiao, 2006).

**6. Enhancing Delegation in Software Management**

Delegation in software management is a critical practice that empowers teams to efficiently tackle complex projects and distribute responsibilities effectively. It involves entrusting specific tasks or decision-making authority to team members based on their expertise and skill sets. Effective delegation in software management should begin with a clear understanding of project objectives and team members' strengths and weaknesses. Project managers should identify the right individuals for each task and provide them with a well-defined scope of work, timelines, and expectations. Regular communication and feedback loops are essential to monitor progress and ensure alignment with project goals. Delegation not only fosters a sense of ownership and accountability among team members but also allows managers to focus on higher-level strategic planning, ultimately contributing to the overall success of software development projects. Nevertrheless, improved project efficiency is a cornerstone of effective delegation in software management. When tasks are strategically distributed to team members based on their qualifications and expertise, it ensures that the right people are handling the right responsibilities. This not only minimizes bottlenecks but also fosters a smoother workflow throughout the project lifecycle. By harnessing the strengths of individual team members, delegation optimizes resource allocation, prevents overburdening of any single team member, and enables a more streamlined execution of tasks. Consequently, project efficiency is significantly enhanced as team members are empowered to work on tasks that align with their skills and knowledge, leading to faster delivery, reduced rework, and ultimately, the successful completion of software development projects within time and budget constraints.

Enhancing project efficiency through effective delegation in software management requires a thoughtful approach. The following points help achieving more effective delegation in software managemen:

- Task Analysis and Matching: Begin by conducting a thorough analysis of project tasks and their requirements. Then, match these tasks to team members based on their skills, experience, and domain expertise. Academic research can help identify optimal task-team member pairings.
- Clear Communication: Emphasize the importance of clear and precise communication when delegating tasks. Academic studies on communication strategies can inform how to articulate expectations, deliver instructions, and establish channels for feedback.
- Delegation Criteria: Develop criteria and guidelines for delegation that take into account task complexity, criticality, and the capabilities of team members. Research can help determine the factors to consider when making delegation decisions.
- Empowerment and Accountability: Encourage a culture of empowerment and accountability within the team. Academic literature on leadership and team dynamics can provide insights into fostering a sense of ownership among team members.
- Monitoring and Feedback: Establish mechanisms for monitoring task progress and providing regular feedback. Research on performance evaluation and feedback loops can guide the design of effective monitoring systems.
- Risk Management: Incorporate risk assessment into delegation decisions. Academic studies on risk management in software projects can help identify potential risks associated with task assignments and how to mitigate them.
- Training and Skill Development: Support ongoing training and skill development for team members. Academic resources on professional development and training methodologies can inform strategies for enhancing team members' capabilities.
- Delegation Tools and Technologies: Explore academic literature on project management tools and technologies that can facilitate delegation and task tracking. Leveraging the right software tools can streamline the delegation process.
- Performance Metrics: Define and measure key performance metrics related to project efficiency and delegation effectiveness. Academic research on project metrics can provide guidance on what to measure and how to interpret results. Adaptation and Continuous Improvement: Encourage a culture of adaptation and continuous improvement in delegation practices. Academic studies on agile methodologies and iterative project management can offer insights into adapting to changing project needs.

**7. Challenges and Best Practices**

This segment will delve into the difficulties that frequently arise when it comes to delegation and leadership, in the software industry. Additionally, it will offer insights, into the approaches that can alleviate these challenges and promote successful software management.

7.1 Challenges in Delegation and Leadership

Difficulties, with delegation and leadership, can have an impact, on the effectiveness and accomplishments of software management. The following Table 6 provides a summary of the challenges. How they can affect outcomes (Overton & Lowry, 2013).

**Table** 6. Challenges in Delegation and Leadership.

| Challenge | Impact |
|---|---|
| Resistance to Change | Hinders adoption of delegation and trust. |
| Trust and Confidence | This may lead to hesitation in delegating tasks. |
| Monitoring and Evaluation | Complicates assessment of delegation effectiveness. |
| Poor Communication | Resulting in misunderstandings and conflicts. |
| Conflicts of Interest | Divergent priorities impede seamless delegation. |
| Inadequate Skill Sets | Undermines task execution and delegation success. |
| Lack of Support and Resources | Impedes effective delegation efforts. |

7.2 Best Practices for Effective Delegation and Leadership

Effective delegation and leadership plays a role as shown in Table 7. By implementing the following practices software managers can overcome challenges (Nordbäck & Espinosa, 2019).

**Table** 7. Best Practices for Delegation and Leadership.

| Best Practice | Description |
|---|---|
| Clear Goal setting | Comprehensive delineation of project objectives. |
| Fostering Trust | Consistent support, mentorship, and guidance to team members. |
| Effective Communication | Open and frequent exchanges between managers and teams. |
| Task-Skill Alignment | Matching responsibilities with individual expertise. |
| Setting Realistic Expectations | Unambiguous guidelines, standards, and quality expectations. |
| Provision of Resources and Support | Adequate training, tools, and information access for teams. |
| Regular Task Progress Monitoring | Constructive feedback and performance assessment. |
| Outcome Evaluation | Continual assessment and process optimization. |

By incorporating these recommended techniques software managers can strengthen their abilities to delegate tasks and lead effectively resulting in enhanced project efficiency increased productivity and successful software management.

## 8. Beneficiaries and Implications

Delegation plays a role in the software industry leading to advantages and important consequences.

8.1 Beneficiaries of Delegation

Both individual programmers and entire organizations benefit from the practice of delegation. For individuals' delegation provides opportunities to develop skills grow personally and find the most satisfactory jobs. It enables them to take on challenges acquire abilities and contribute to innovative solutions. At this level, delegation promotes increased efficiency, collaboration, innovation, and risk management. This leads to engaged teams ultimately resulting in higher productivity and successful project outcomes (Steffel , Williams, & Perrmann-Graham, 2016).

8.2 Implications for Software Management

The significance of delegation and leadership, in software management reaches far and wide. It highlights the value of communication, building trust, and embracing transformational leadership approaches. These actions play a role in enhancing project efficiency, team productivity, and overall organizational achievements. However, it's important to acknowledge that delegation should be customized for every software project while considering the requirements and dynamics of the team (F.A.Z & R.D.F.S.M, 2020).

## 9. Conclusion

In this concluding section, it will present the outcomes of this study, summarize the insights, and outline directions, for future research. It will also discuss the implications of our research for both understanding and practical application. This study delved into the intricacies of delegation and leadership within software management. Notably, the findings highlighted that effective delegation is crucial in enabling managers to focus on higher-level tasks, while also fostering the growth and development of team members. This success heavily relies on communication, transparent expectations, and building trust between managers and their teams. Among leadership styles examined transformational leadership is also essential, which are characterized by inspiration and guidance. Emerged as an effective approach in software management. Additionally, organizational culture was found to play a role in shaping how delegation and leadership practices are adopted. The implications of this research carry importance for both understanding and practical implementation. Firstly, the study underscores the role played by delegation in enhancing efficiency and nurturing team growth. Clear communication and trust building are factors ensuring delegation within organizations. Furthermore, the findings highlight that transformational leadership is highly preferred in software management emphasizing the importance of inspiring others while providing guidance. Despite these insights gained from this research efforts, there remains a gap for further exploration, within this domain. To gain an understanding for future study, should explore how cultural and organizational factors, long-term investigations and different leadership approaches affect the management of software. It is also important to investigate the effects of delegation and leadership, on the end users' satisfaction and their overall software experience. In conclusion, this study contributes to the intricacies of delegation and leadership, in software management this comprehensive guide provides insights to enhance and optimize software management practices in a manner. By embracing effective delegation and transformational leadership, organizations can enhance project efficiency, team productivity, and overall success.

## 10. Future Work

Future work regarding this paper offers an exciting avenue for in-depth exploration of each of the the points mentioned in section 6. Firstly, a deeper investigation into task analysis and matching can involve developing sophisticated algorithms or AI-based systems to automate the task allocation process, taking into account not only skills but also team members' preferences and workload. Secondly, the realm of clear communication could benefit from studies on the impact of different communication mediums and tools, as well as the cultural aspects that influence effective communication in diverse software development teams. Thirdly, the establishment of delegation criteria could be refined through statistical models that consider historical project data and its correlation with delegation outcomes. Moreover, empirical studies might unveil the nuances of empowerment and accountability, shedding light on the psychological factors that influence team members' sense of ownership. Additionally, research could delve into innovative ways of monitoring and feedback using advanced analytics and natural language processing techniques. Risk management might benefit from predictive analytics models that forecast potential risks based on historical patterns and project attributes. The exploration of training and skill development could involve designing personalized learning paths based on individual team members' learning styles and preferences, leveraging insights from educational psychology. For delegation tools and technologies, future work may involve the development of AI-driven tools that not only assist in task allocation but also facilitate real-time resource allocation and optimization. The definition and measurement of performance metrics could be fine-tuned through comprehensive benchmarking studies, while adaptation and continuous improvement could be informed by agile maturity models and organizational change management theories. These directions for future research would contribute to a more comprehensive understanding of effective delegation in software management and provide actionable insights for practitioners aiming to enhance project efficiency.